\documentclass[%
 reprint,
 amsmath,amssymb,
 aps,showkeys
]{revtex4-2}

\usepackage{graphicx}
\usepackage{dcolumn}
\usepackage{bm}
\usepackage{amsmath}
\usepackage[utf8]{inputenc} 
\usepackage[T1]{fontenc}
\usepackage{tabularx}
\usepackage{gensymb}
\usepackage{xcolor}
\usepackage{svg}
\usepackage{ragged2e}

\begin{document}

\preprint{APS/123-QED}

\title{Correlated inhomogeneous absorption profiles across distinct optical transitions in a rare-earth doped crystal}


\author{Flora Segur$^{1,2}$}
\author{Sacha Welinski$^2$}
\author{Alban Ferrier$^3$}
\author{Perrine Berger$^2$}
\author{Anne Louchet-Chauvet$^1$}

\affiliation{$^1$Institut Langevin, CNRS, ESPCI Paris, Université PSL, Paris, France}
\affiliation{$^2$Thales Research and Technology, Palaiseau, France}
\affiliation{$^3$Institut de Recherche de Chimie de Paris, CNRS, Université Paris Sciences et Lettres,\\ Sorbonne Université, Faculté des Sciences et Ingénierie, UFR 933, Paris, France}

\begin{abstract}
In rare-earth ion-doped crystals, inhomogeneous absorption profiles reflect the distribution of local environments experienced by individual ions. While each optical transition probes this distribution differently, their fine spectral structures may retain correlations arising from shared local perturbations. In this paper, we present a low-temperature, high-resolution spectroscopic study in 
$\mathrm{Er^{3+}\!:YSO}$  of the transition
$\mathrm{^4I_{15/2} \rightarrow\ ^4I_{11/2}}$  at $980$~nm and compare it to the well-known transition
$\mathrm{^4I_{15/2} \rightarrow\ ^4I_{13/2}}$  at 1.5~$\mu$m. Using spectral hole burning on one transition while monitoring the other, we uncover for the first time spectral correlations between two optical transitions, providing new insight into the microscopic origin of inhomogeneous distributions in rare-earth-doped crystals.

\end{abstract}

\maketitle

\section{\label{sec:introduction}Introduction}

Trivalent rare earth ion (REI) doped crystals exhibit the unique property of extremely narrow homogeneous linewidths in some of their 4f-4f transitions at cryogenic temperatures \cite{equall_ultraslow_1994}. The ratio between the inhomogeneous $\Gamma_{inh}$ and homogeneous $\Gamma_h$ linewidths can be exceptionally large ($> 10^7$), enabling numerous applications in both classical and quantum information processing using spectral hole burning (SHB) techniques \cite{cone_rare-earth-doped_2012,putz_spectral_2016,boettger_material_2003}. Until now, SHB has been employed primarily between two optically-coupled 4f states of REIs. Recently SHB has also been observed on electronic spin transitions, using narrowband radiofrequency excitation \cite{putz_spectral_2016}. A particularly interesting property would be the ability to engrave spectral features on one optical absorption line, thereby inducing similar features on another absorption line. Such a phenomenon would benefit many applications, including radiofrequency spectral analysers, quantum memories, and, more generally applications involving photons at different wavelengths.
In this article, we present the evidence of correlations in the inhomogeneous absorption profiles of two 4f–4f transitions in a single REI species. Our system is $\mathrm{Er^{3+}}$-doped $\mathrm{Y_2SiO_5}$ (YSO) crystal, widely recognized for its narrow and highly coherent optical transitions $\mathrm{^4I_{15/2} \leftrightarrow\ ^4I_{13/2}}$ which are compatible with low-loss telecom fibers at 1.5~$\mu$m \cite{boettger_material_2003,louchet-chauvet_telecom_2020}. $\mathrm{Er^{3+}}$ also presents transitions $\mathrm{^4I_{15/2} \leftrightarrow\ ^4I_{11/2}}$ near 980~nm, a convenient wavelength for narrow lasers and external modulators. However, the properties of these lines at low temperature have not previously been established by high-resolution spectroscopy. The first part of this study therefore focuses on the characterization of the 980~nm transitions, in order to compare their properties with those of the 1.5~$\mu$m transitions. To this end, we investigated the spectral properties through polarization-dependent absorption and the effective $g$-factor. We also measured the dynamical properties to evaluate the population lifetime $T_1$ and the coherence time $T_2$. The second part of the study focuses on the experimental investigation of correlations between 1.5~$\mu$m and 980~nm transitions. By optically pumping one transition and probing the other, we observed an induced spectral hole, which we subsequently characterized. By shifting the position of the pumped hole, we observed that the properties of the induced hole vary across the inhomogeneous profile of the readout transition. By pumping at 1.5~$\mu$m or at 980~nm, we observed distinct behaviors that reveal how ions are spectrally distributed between the two transitions. Finally, we used a simple model to statistically explain these observations. These results offer new insight into the role of the local disorder, and suggest potential for multi-color architectures in classical and quantum information processing.

\section{\label{sec:materialbackground} $\mathrm{Er^{3+}:Y_2SiO_5}$ material background}

The monoclinic yttrium orthosilicate ($\mathrm{Y_2SiO_5}$, YSO) crystal belongs to the low symmetry space group $C_{2h}^6$, with two inequivalent Y$^{3+}$ sites described by $C_1$ symmetry. When inserted into the matrix, the $\mathrm{Er^{3+}}$ ions substitute to Y$^{3+}$ ions without charge compensation \cite{scientific-materials_y2sio5_nodate}. Since erbium is a Kramers ion, the crystal field lifts the degeneracy of its free-ion levels $\mathrm{^{2S+1}L_{J}}$, splitting each J manifold into $N=(2 J+1)/2$ doublet sublevels, which are labeled $Z_1,…,Z_N$ for the ground state manifold. The energy diagram is presented in Fig.~\ref{fig:fig1}(a). The three-level system formed by $\mathrm{^4I_{15/2}}$ (Z), $\mathrm{^4I_{13/2}} $ (Y), and $\mathrm{^4I_{11/2}} $ (X) has initially been studied in the context of solid-state lasers \cite{laporta_single-mode_1993,li_spectroscopic_1992}. In the literature, the two excited states X and Y were investigated in 1\% $\mathrm{Er^{3+}:YSO}$ at room temperature \cite{li_spectroscopic_1992}. The fluorescence decay times of the two manifolds $Y$ and $X$ were measured to be 8~ms and 18~$\mu$s respectively, the latter attributed to fast non-radiative multiphonon relaxation to the lower-lying Y levels. Indeed, the phonon cutoff frequency is quite high in YSO (960 cm$^{-1}$) \cite{layne_multiphonon_1977,campos_spectroscopic_2004} which favors the occurrence of multiphonon radiative relaxation along the 980~nm transition. 
The optical branching ratios from the $\mathrm{^4I_{11/2}}$ level were determined to be 0.17 ($X \rightarrow Y$) and 0.83 ($X \rightarrow Z$) \cite{li_spectroscopic_1992}. \\

At low temperature, the $Z_1 \leftrightarrow Y_1$ transition has been extensively characterized in weakly concentrated samples (typically between 10 and 50 ppm) \cite{bottger_spectroscopy_2006,sun_magnetic_2008,bottger_optical_2006}. The two substitution sites for $\mathrm{Er^{3+}}$ ions within the YSO crystal matrix give rise to two absorption lines in the telecom C-band: at 1536.48~nm for site 1 and 1538.90~nm for site 2, with respective fluorescence lifetimes of 11.4~ms and 9.2~ms at 3.5~K \cite{bottger_laser_2002}. Homogeneous linewidths down to 73 Hz were measured under 7 T magnetic field \cite{bottger_effects_2009}, raising the interest of this telecom transition for quantum applications. 
In contrast, the $Z_1 \leftrightarrow X_1$ transition remains poorly characterized. More specifically, no low-temperature, high-resolution spectroscopic studies of this transition have been reported in crystalline systems, including YSO. To our knowledge, the only high-resolution spectroscopy of this transition was conducted in an erbium-doped fiber, an amorphous medium \cite{kamel_erbium-doped_2025}.

\section{\label{sec:experimentalsetup} Experimental setup}
\begin{figure}[h]
\includegraphics[width=0.99\linewidth]{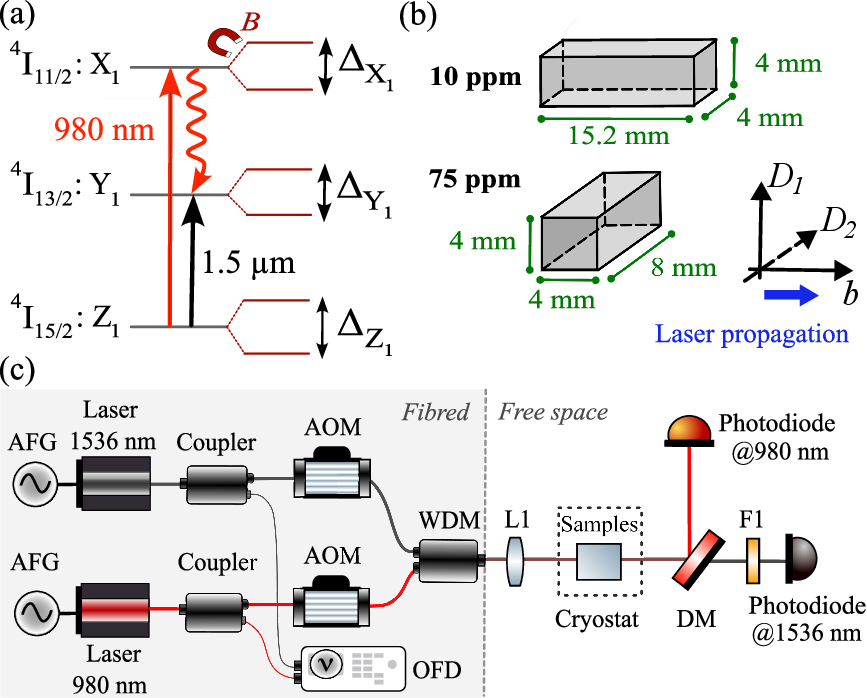}
\caption{\label{fig:fig1} (a) Simplified state diagram of Er$^{3+}$:YSO and the corresponding transitions $Z_1 \leftrightarrow Y_1$ and $Z_1 \leftrightarrow X_1$. (b) Crystals dimensions along the dielectric axes $D_1$, $D_2$ and the crystalline \textit{b}-axis. (c) Experimental setup: the left side is fibered and the right side is free space.}
\end{figure}

We use two Er$^{3+}$:YSO crystals with 10~ppm and 75~ppm atomic concentrations. The dimensions of each crystal along the dielectric axes $D_1$, $D_2$ and $b$ are shown in Fig.~\ref{fig:fig1}(b). The light propagates along the $b$ axis. The experimental setup is illustrated in Fig.~\ref{fig:fig1}(c). For the optical excitation of the $Z_1 \leftrightarrow Y_1$ transition, we have two laser sources at our disposal. For monochromatic excitation, we use a Koheras X15 fiber laser (<0.1~MHz linewidth), providing 12 mW at 1536.48~nm on the sample. For GHz-range frequency chirps, we use a Gooch\&Housego DFB laser (up to 35 mW on sample), whose frequency is controlled through current modulation using an arbitrary function generator (AFG). The $Z_1 \leftrightarrow X_1$ transition is addressed using a Newport Velocity TLB-6700 tunable external cavity diode laser (<~0.2~MHz Lorentzian linewidth, 1.8~mW on sample), which is frequency-stabilized when not scanned. The relative instantaneous frequency of the lasers is measured in real time using an optical frequency discriminator (OFD) (Silentsys). Pulses and small frequency scans (up to 50~MHz) are obtained with fibered Acousto-Optic-Modulators (AOM). The two beams are combined with a wavelength demultiplexer (WDM, model WP9850A) and collimated by a lens (L1) with a 4.6~mm focal length, resulting in a collimated beam (diameter approximately of 850~$\mu$m at 980 nm) which is sent through the sample. The latter sits on a vibration-damped copper mount cooled to 3.5~K by thermal links to the cold finger of a closed-cycle cryostat (Montana Instruments s50 Cryostation). The two wavelengths are separated at the cryostat output using a dichroic mirror (DM). A reflective filter (F1) is subsequently placed in the 1.5~$\mu$m beam to remove any residual 980~nm component. The light at 1536~nm is detected with an InGaAs photodiodes (PDA20C2, or PDB230C in the case of photon echo measurements) and the light at 980~nm using a Si avalanche photodetector (APD 110A/M). For photon-echo experiments, an additional 50~mm focal lens is positioned after L1 to increase the beam intensity $I$ inside the crystal, thus enhancing the Rabi frequency $\Omega_{in}\propto \sqrt{I}$, with a resulting beam waist of approximately 29~$\mu$m at 980~nm.

\section{\label{sec:spectro}HIGH RESOLUTION SPECTROSCOPY OF THE $Z_1 \leftrightarrow X_1$ TRANSITION }
\subsection{\label{sec:absorption}Absorption and polarization dependence}

\begin{figure}[t!]
\includegraphics[width=0.8\linewidth]{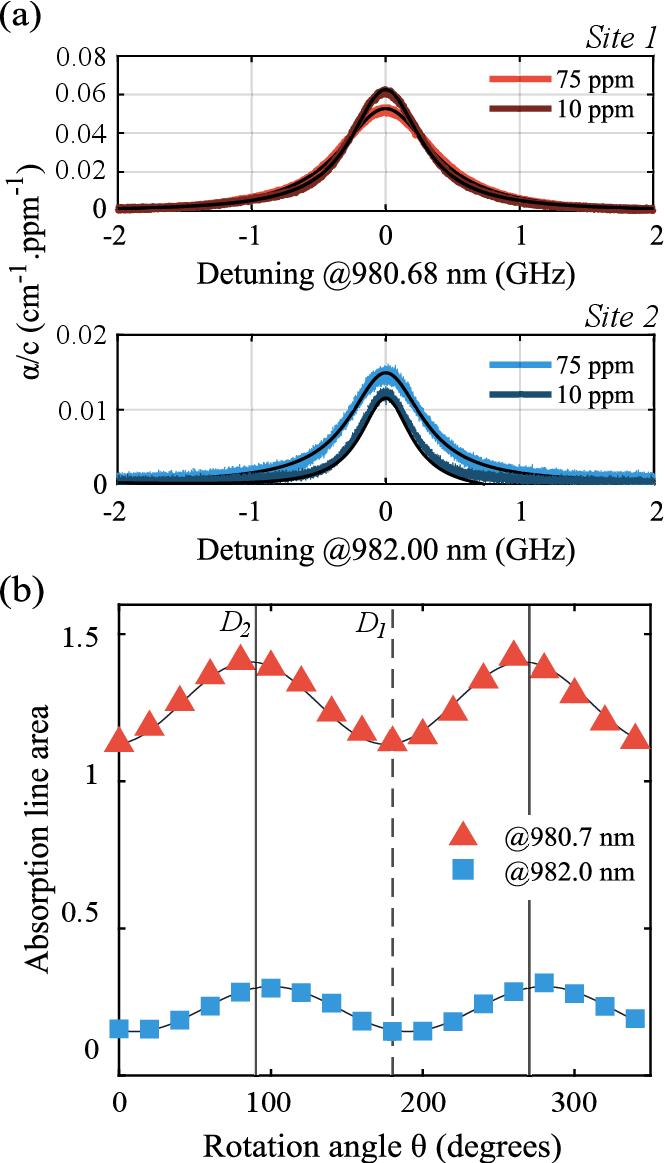}
\caption{\label{fig:fig2}(a) Absorption profiles of the $Z_1 \leftrightarrow X_1$ transition for the two sites of Er$^{3+}$:YSO (resonant at 980~nm and 982~nm), measured for 10~ppm and 75~ppm doping concentration. The polarisation is along the $D_2$ axis. The site assignation is explained in Section~\ref{sec:siteassign}. (b) Absorption line area of $Z_1 \leftrightarrow X_1$ in 75~ppm-Er$^{3+}$:YSO versus polarization angle $\theta$ in the ($D_1,D_2$) plane, for site 1 (red triangles) and site 2 (blue squares). Black lines correspond to adjustments to equation (1).  }
\end{figure}

We probe the absorption profiles of both $\mathrm{Er^{3+}:YSO}$ samples (10 and 75~ppm) in the 980~nm range using a 0.5~ms-long, quasi linear optical chirp spanning 5~GHz. Two absorption lines are visible, see Fig.~\ref{fig:fig2}(a), corresponding to the two sites of erbium ions, centered at 980.68~nm in vacuum (10197.00 $\mathrm{cm}^{-1}$) and the other at 982.00~nm in vacuum (10183.30 $\mathrm{cm}^{-1}$). 
All lines are well fitted by a Lorentzian profile, as it is the case for the $Z_1 \leftrightarrow Y_1$ lines (see Appendix A). The respective FWHM of the lines $Z_1 \leftrightarrow Y_1$ and $Z_1 \leftrightarrow X_1$ 
are reported in Table~\ref{tab:table1}. The measured linewidths at 980.68~nm are very similar to those at 1536.48~nm, which is close to previously reported values \cite{bottger_spectroscopy_2006}. We also observe a slight dependence of the linewidth with respect to the concentration. At such a low doping concentration, it is commonly observed that the REI concentration marginally contributes to the local disorder, which is dominated by defects in the host matrix \cite{konz_temperature_2003,FERRIER2016406}. The optical absorption maximum in the $Z_1 \leftrightarrow X_1$ transition at 980.68~nm is weaker than that of the $Z_1 \leftrightarrow Y_1$, suggesting a lower oscillator strength.

\begin{table}[h!]
\begin{ruledtabular}
\begin{tabular}{cl|c|c}
\multicolumn{2}{c|}{Crystal concentration} &
10~ppm &
75~ppm \\
\colrule
\hline
\multicolumn{2}{c|}{Transitions} &
\multicolumn{2}{c}{FWHM (GHz)} \\
\colrule
$Z_1 \leftrightarrow Y_1$&1536.48~nm & 0.59 & $0.79$ \\
$Z_1 \leftrightarrow X_1$ & 980.68~nm & 0.64 & $0.85$ \\
 & 982.00~nm & 0.46 & $0.67$ \\
\end{tabular}
\caption{\label{tab:table1} Measured inhomogeneous linewidth (FWHM) of the absorption line of the $Z_1 \leftrightarrow Y_1$ and $Z_1 \leftrightarrow X_1$ transition in both crystals.}
\end{ruledtabular}
\end{table}

Furthermore, we investigate the polarization dependence of the $Z_1 \leftrightarrow X_1$ absorption line in the 75~ppm-doped crystal, by rotating the laser polarization axis $\mathbf{E}$ within the ($D_1$,$D_2$) plane, keeping the propagation axis along $b$. As shown in Fig.~\ref{fig:fig2}(b), the absorption line area is computed for each angle $\theta$ between $D_1$ and $\mathbf{E}$. For both sites, we find that the absorption varies sinusoidally with the laser polarization, with a maximum close to $D_2$. This was also observed for the $Z_1 \leftrightarrow Y_1$ transition which is, unlike the only-electric-dipole-allowed transition $Z_1 \leftrightarrow X_1$, a hybrid- electric-magnetic-dipole transition \cite{petit_demonstration_2020}. Following \cite{dinndorf_principal_1992}, one can adjust the data points with the function that describes the absorption cross-section:
\begin{equation}
d(\theta) = \frac{d_1 + d_2}{2} + \frac{d_1 - d_2}{2} \cos(\theta - \theta_0)
\end{equation}

We obtain $\theta_0=-4.0(1.1)\degree $ for site 1 and $\theta_0=10.0(0.7)\degree$ for site 2. Please note that the uncertainties are explicitly stated in parentheses in the whole article. The complete fit parameters are given in Appendix B.

\subsection{\label{sec:siteassign}Site assignation}
\begin{figure}[t]
\includegraphics[width=0.8\linewidth]{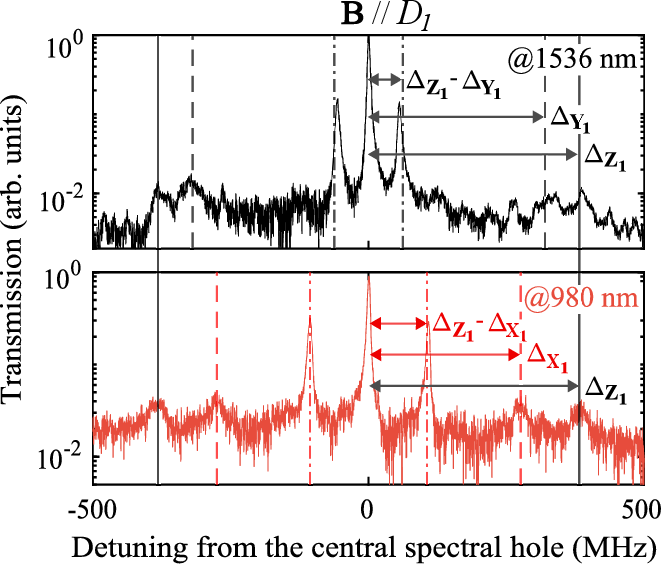}
\caption{\label{fig:fig3} Spectral hole burning transmission spectra of site 1 in 75~ppm-Er$^{3+}$:YSO under a $\sim$ 10 mT magnetic field along $D_1$. The upper panel shows the $Z_1 \leftrightarrow Y_1$ transition, and the lower panel shows the $Z_1 \leftrightarrow X_1$ transition. }
\end{figure}

In order to ascribe each line (980.68~nm and 982.00~nm) to a given site, we study the side features of hole burning spectra under magnetic field in the 75~ppm sample. A magnetic field $|\mathbf{B}|$ of approximately 10~mT is applied along the $D_1$ axis with a NdFeB permanent magnet placed outside the sample chamber. A spectral hole is burnt sequentially at 980.68~nm, 982.00~nm and 1536.48~nm, with the latter corresponding to site 1 \cite{sun_magnetic_2008}. Transmission spectra are acquired by scanning the laser frequency over 1~GHz around the hole frequency, 1.5~ms after the burning pulse (see Fig.~\ref{fig:fig3}). Under our magnetic field and temperature conditions, the spin-lattice relaxation rapidly redistributes the atomic populations within the ground state Zeeman sublevels, leading to six side holes on each spectrum \cite{hastings-simon_zeeman-level_2008}. The position of these side holes depends on the Zeeman splittings of the ground state $\Delta_{Z_1}$ and the excited states $\Delta_{Y_1}$ and $\Delta_{X_1}$, displayed on Fig. 1 (a). The strong side holes visible on the 1.5~$\mu$m spectrum are found at $\pm | \Delta_{Z_1}-\Delta_{Y_1}|$ of site 1. The smaller side holes are found at $\pm \Delta_{Z_1}$ and $\pm \Delta_{Y_1}$. The holeburning spectra at 980~nm and 982~nm also exhibit 6 side holes located at $\pm | \Delta_{Z_1}-\Delta_{X_1}|$, $\pm \Delta_{Z_1}$ and $\pm \Delta_{X_1}$. We observe that two side holes in the 980~nm spectrum coincide with the $\pm \Delta_{Z_1}$ hole pair observed in the 1.5~$\mu$m holeburning spectrum. 
When $\mathbf{B}//D_1$, the effective ground state $g$-factor is very different for site 1 (5.54(0.07)) and site 2 (15.02(0.02)) \cite{sun_magnetic_2008}. As a result, the $\Delta_{Z_1}$ hole from the other site is located far from the observed holes, ensuring a clear and reliable identification of the site. From this observation we ascribe the optical transition at 980.68~nm to site 1 and the 982.0~nm transition to site 2. 


\subsection{\label{sec:effectivegtensor}Effective $g$-factor}

\begin{figure}[t!]
\includegraphics[width=0.82\linewidth]{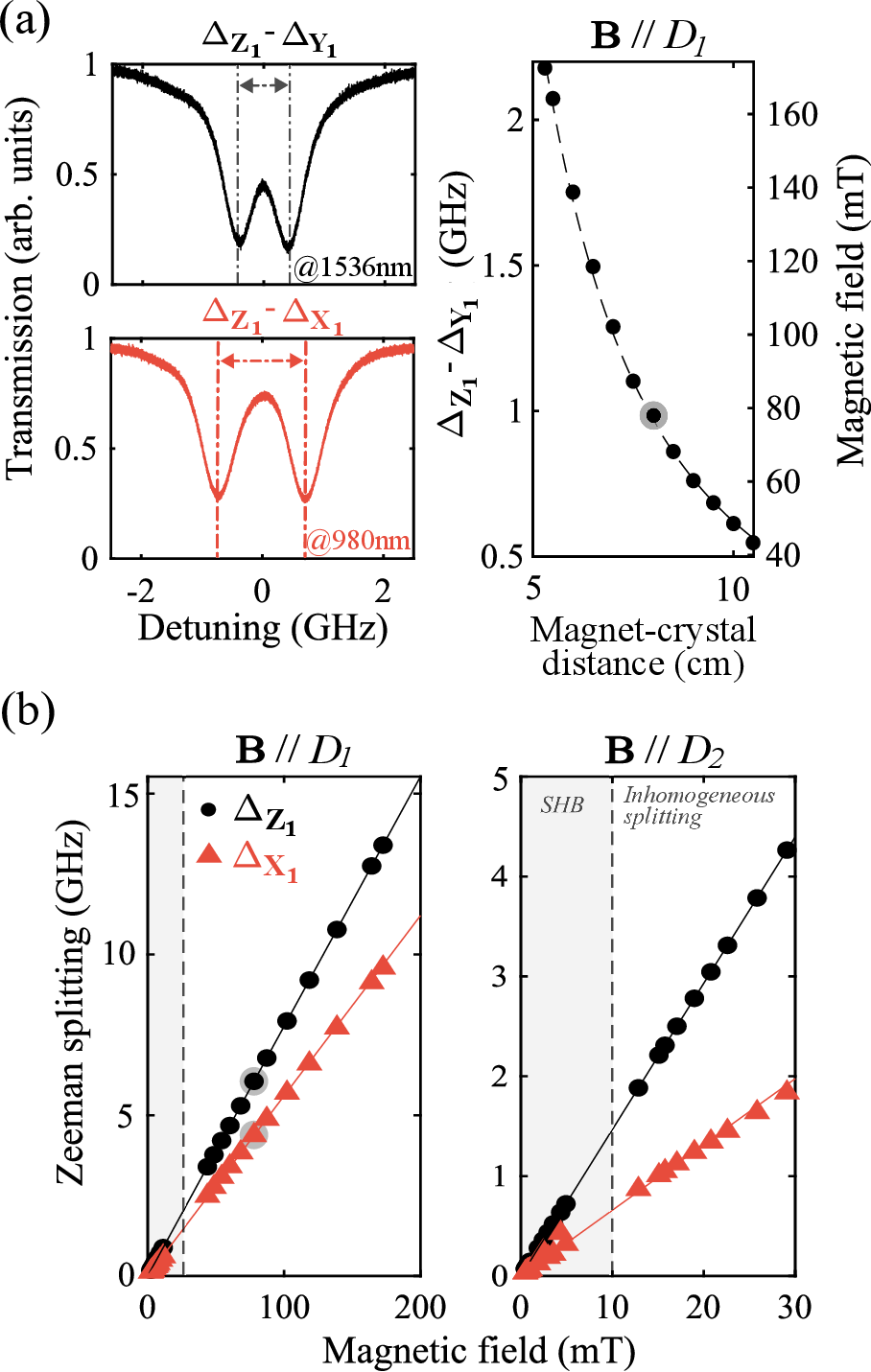}
\caption{\label{fig:fig4} (a) Left panel: Zeeman splitting spectra for site 1 in 10~ppm-Er$^{3+}$:YSO with magnetic field along $D_1$. The $Z_1 \leftrightarrow Y_1$ (top) and $Z_1 \leftrightarrow X_1$ (bottom) transitions are measured for $|\mathbf{B}|=$~79~mT with $\mathbf{B} // D_1$ (black dot in shaded area, right panel). Right panel: magnetic field measured from the $\Delta_{Z_1}-\Delta_{Y_1}$ splitting versus magnet–crystal distance $x$, with a $\propto 1⁄x^2$ fit. (b) Zeeman splitting $\Delta_{Z_1}$ (black circles) and $\Delta_{X_1}$ (red triangles) versus magnetic field for $\mathbf{B} // D_1$ and $\mathbf{B} // D_2$. Dashed line separates data obtained with SHB (Fig.~\ref{fig:fig3}) and inhomogeneous profile measurements (Fig.~\ref{fig:fig4}(a)). The solid lines correspond to linear fits adjusted with respect to $|\mathbf{B}|$. }
\end{figure}

When a magnetic field is applied, the electronic Zeeman splitting in the ground and excited states can be written $\Delta_k=\mu_B g_{k,\mathrm{eff}} |\mathbf{B}|/h$, where $g_{k,\mathrm{eff}}$ denotes the dimensionless effective $g$-factor of the $k$ level ($k=(Z_1,Y_1,X_1)$) along the magnetic field direction with respect to the crystal axes, $\mu_B$ the Bohr magneton and $h$ the Planck constant. Depending on the magnetic field amplitude, two methods are used to evaluate the effective $g$-factor $g_{X_1,\mathrm{eff}}$, from the measurement of $\Delta_{X_1}$ and $|\mathbf{B}|$. At low magnetic fields ($|\mathbf{B}| <$~20~mT) we separately perform spectral hole-burning experiments on the two transitions (see Fig.~\ref{fig:fig3}). We examine the holeburning spectrum at 980~nm and identify 6 side holes, which are expected at positions $\pm \Delta_{Z_1}$, $\pm \Delta_{X_1}$ and $\pm | \Delta_{Z_1}-\Delta_{X_1}|$. Comparing the two spectra, we locate the $\pm \Delta_{Z_1}$ side holes that appear on both spectra and establish that $\Delta_{Z_1}>\Delta_{X_1}$ for $\mathbf{B}//D_1$ and $\mathbf{B}//D_2$ (see Appendix C). Using this information, we are able to infer $\Delta_{X_1}$ from the position of the strongest holes $\pm | \Delta_{Z_1}-\Delta_{X_1}|$. The magnetic field strength is calibrated using the holeburning spectrum on the $Z_1 \leftrightarrow Y_1$ transition. At higher fields (typically 20~mT~$< |\mathbf{B}| <~$200~mT), the inhomogeneously broadened absorption profile is split into two components (see Fig.~\ref{fig:fig4}(a)). The splitting in the $Z_1 \leftrightarrow X_1$ transition provides the value of $\Delta_{Z_1}-\Delta_{X_1}$ for the measured magnetic field. Again, the magnetic field is calibrated using the 1.5~$\mu$m line splitting. We fit $\Delta_{X_1}$ as a linear function of $|\mathbf{B}|$ and extract the effective $g$-factor $g_{X_1,\mathrm{eff}}$. The resulting effective $g$-factors measured along $D_1$ and $D_2$, are given in Table~\ref{tab:table2}.

\begin{table}[h!]
\begin{ruledtabular}
\begin{tabular}{rlcc}
\multicolumn{2}{c}{Magnetic field direction} &
\multicolumn{1}{c}{$D_1$} &
$D_2$ \\
\colrule
\hline
Site 1&$g_{Z_1,\mathrm{eff}}$ \cite{sun_magnetic_2008} & 5.54(0.07) & 10.46(0.05) \\
 & $g_{Y_1,\mathrm{eff}}$ \cite{sun_magnetic_2008} & 4.64(0.13) & 6.90(0.09) \\
 & $g_{X_1,\mathrm{eff}} $ (this work) & 3.981(0.009) & 4.58(0.07) \\
\end{tabular}
\caption{\label{tab:table2} Effective $g$-factors for the $Z_1$, $Y_1$ and $X_1$ states, for a magnetic field applied along $D_1$ or $D_2$. The values for $X_1$ are this work’s contribution. }
\end{ruledtabular}
\end{table}



\subsection{\label{sec:holelifetime}Hole lifetime}
\begin{figure}[t]
\includegraphics[width=0.8\linewidth]{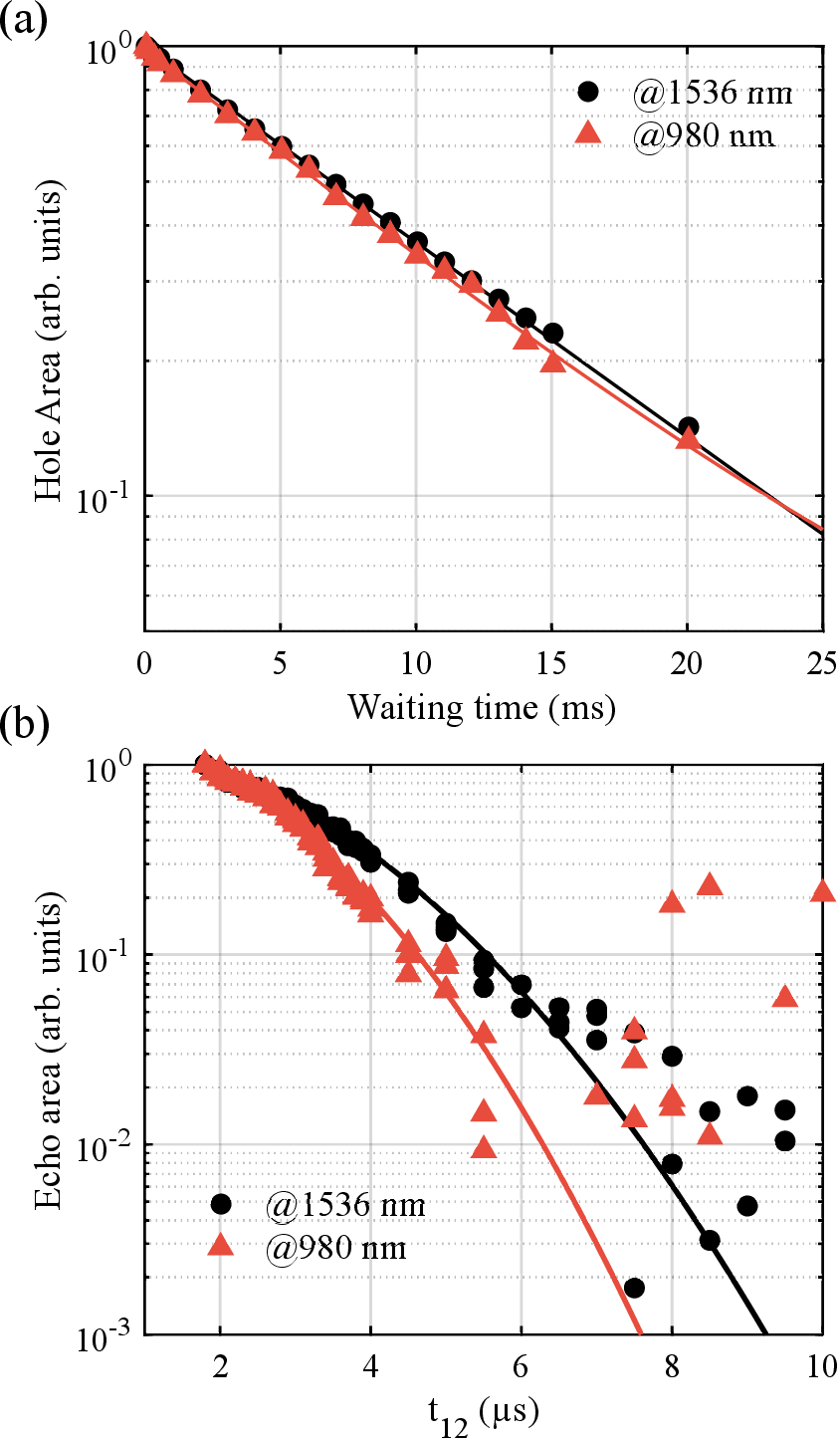}
\caption{\label{fig:fig5} (a) Hole area versus waiting time in 75~ppm-Er$^{3+}$:YSO for site 1 ($Z_1 \leftrightarrow Y_1$, $Z_1 \leftrightarrow X_1$), zero magnetic field. The solid lines are exponential fits to the data, and include a small background. (b) Two-pulse photon echo amplitude in 10~ppm-Er$^{3+}$:YSO, plotted as a function of the delay $t_{12}$, for $Z_1 \leftrightarrow Y_1$ (black circles) and $Z_1 \leftrightarrow X_1$ (red triangles). The solid lines are the adjusted Mims function $E(t)$. }
\end{figure}

In Section~\ref{sec:siteassign}, we showed that SHB can be achieved on the $Z_1 \leftrightarrow X_1$ transition at 3.5 K. We now investigate the spectral hole lifetime by optically pumping the ions during 10~ms, and reading out the spectral hole with a 100~$\mu$s pulse scanned over a 50~MHz frequency range, after a variable delay. We perform this operation separately on the $Z_1 \leftrightarrow Y_1$ and $Z_1 \leftrightarrow X_1$ transitions of site 1, and obtain hole lifetimes of $10.00(0.02)$~ms and $9.1(0.1)$~ms, respectively (see Fig.~\ref{fig:fig5}(a)). These two values are very close because in practice the population excited into $X_1$ accumulates in the long-lived state $Y_1$ by optical pumping \cite{li_spectroscopic_1992}. The hole lifetime that we obtain on the $Z_1 \leftrightarrow Y_1$ transition is in good agreement with previously reported values \cite{bottger_laser_2002}. Unfortunately, given the experimental sequence constrains we can only infer that the $X_1$ population lifetime is shorter than 50~$\mu$s and cannot confirm the 19~$\mu$s value reported in \cite{li_spectroscopic_1992} at 12~K.

\subsection{\label{sec:level2}Optical coherence lifetime}
In order to evaluate the optical decoherence of both transitions $Z_1 \leftrightarrow X_1$ and $Z_1 \leftrightarrow Y_1$ of site 1, we perform 2-pulse photon echo sequence in the 10~ppm crystal, with no applied magnetic field. The excitation pulses (0.9~$\mu$s each) are separated by a variable delay $t_{12}$ and the echo is detected using heterodyne detection to boost the measurement sensitivity \cite{beaudoux_emission_2011}. We measure the photon echo amplitude area as a function of $t_{12}$ and fit its decay with the Mims function $E(t) = E_0 \exp\left( -\left(\frac{2t}{T_M}\right)^x \right)$ [23] with $T_M$ the echo phase memory time. The data and corresponding fits are shown on Fig 5(b). We obtain:
\begin{align*}
  T_M &= 6.9(0.2)\: \mu\text{s} \quad \text{and} \quad x = 2 \quad \text{for } Z_1 \leftrightarrow Y_1 \\
  T_M &= 5.6(0.4)\: \mu\text{s} \quad \text{and} \quad x = 2 \quad \text{for } Z_1 \leftrightarrow X_1
\end{align*}
The non-exponential decay ($x=2$) reflects the spectral diffusion due to magnetic dipole-dipole interaction between $\mathrm{Er^{3+}}$ ions \cite{bottger_optical_2006}. Our results can be compared to those reported \cite{macfarlane_measurement_1997}, where a shorter decay time of 3.7~$\mu$s and $x=1$ were measured in a 32~ppm-doped crystal at 1.5~K.
These results indicate that the decoherence mechanisms affecting each transition are of similar magnitude, given the close values of $T_M$ obtained for both transitions. 

\section{\label{sec:level1}CORRELATIONS ACROSS THE INHOMOGENEOUS DISTRIBUTIONS }
In this section, we examine correlations across the inhomogeneous distributions of transitions $Z_1 \leftrightarrow Y_1$ and $Z_1 \leftrightarrow X_1$. To that end, we perform SHB on one transition and probe the other while the spectral hole survives. All the experiments described in the following are achieved in the 75~ppm-doped crystal (unless otherwise specified in the figure caption), with a laser polarization along $D_2$.

\subsection{\label{sec:pumping1536}Pumping at 1536~nm, readout at 980~nm}

\begin{figure}[t!]
\centering
\includegraphics[width=0.82\linewidth]{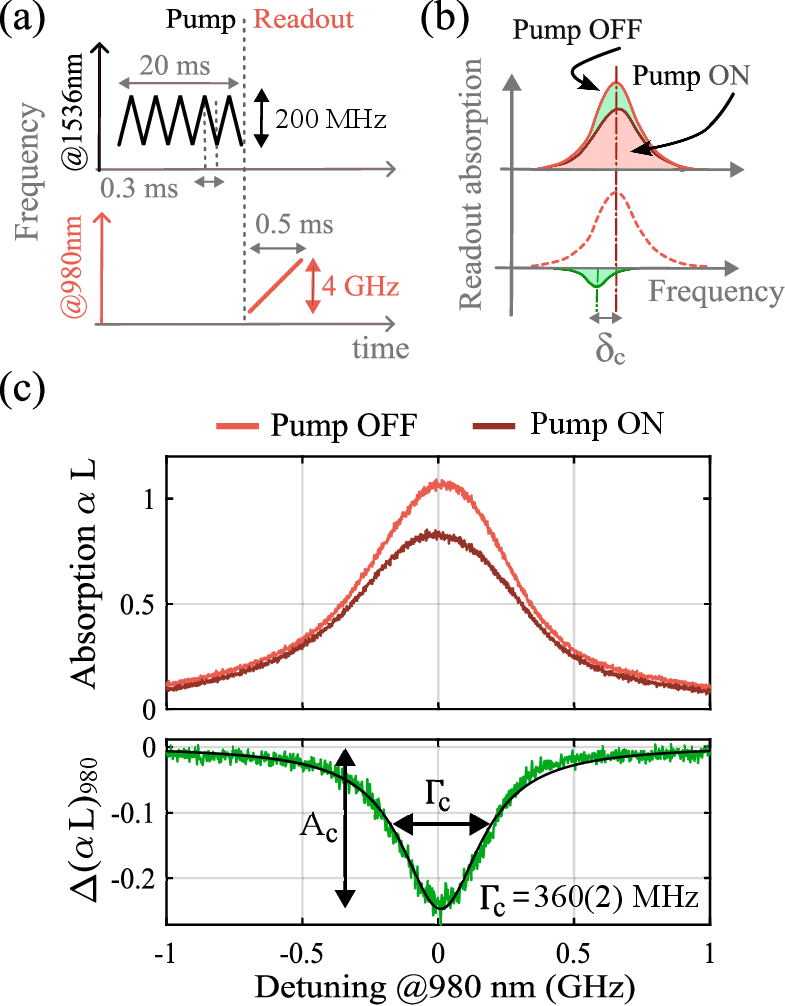}
\caption{\label{fig:fig6} (a) Experimental chronogram. (b) Illustration of absorption spectra of the probed inhomogeneous profile with and without the pump beam (top), highlighting the induced spectral hole detuned by $\delta_c$ (bottom, green). (c) (top) Absorption spectra of the 980~nm transition for a hole detuning at 1536~nm close to zero, in 10~ppm-Er$^{3+}$:YSO. (bottom) Resulting absorption modification $\Delta (\alpha L)=\alpha L(\nu)_{on}-\alpha L(\nu)_{off}$. The fitted parameters are $A_c=-0.248(0.001)$ and $\delta_c=-9.0(0.4)$~MHz.}
\end{figure}

\begin{figure}[t!]
\includegraphics[width=0.8\linewidth]{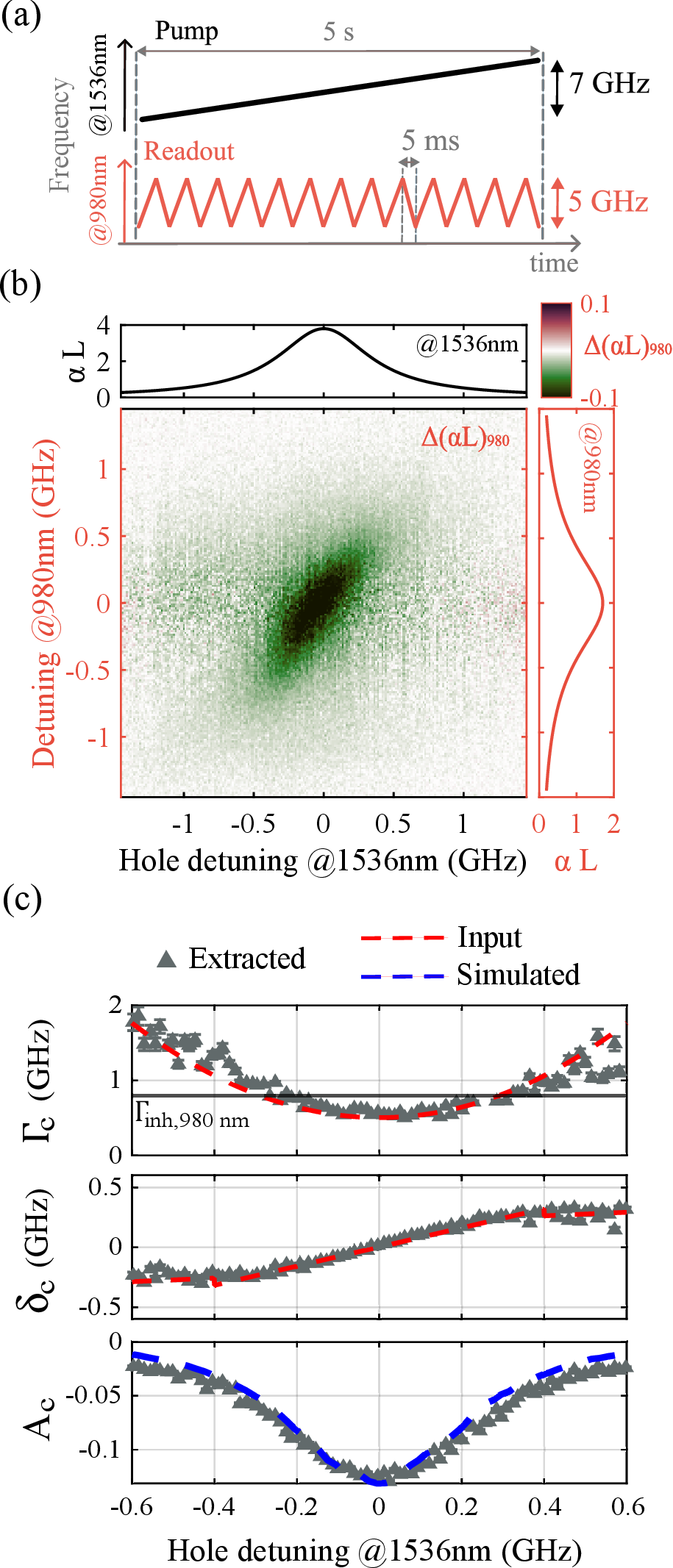}
\caption{\label{fig:fig7} (a) Experiment chronogram. (b) Colorplot of the induced spectral hole $\Delta (\alpha L)$ at 980~nm versus hole detuning at 1536~nm with the corresponding inhomogeneous profile shown in the side panel, for 75~ppm-Er$^{3+}$:YSO. (c) Grey triangles: fitted parameters (linewidth $\Gamma_c$, amplitude $A_c$, and position $\delta_c$) of $\Delta (\alpha L)$ as a function of hole detuning at 1536~nm. Red dashed lines: (top) parabolic fit to the data; (middle) piecewise linear fit to the data with slope 0.8 for hole detuning~$\in [-0.4, 0.4]$. These parameters are used as input parameters for the simple model of Section~\ref{sec:simplemodel}. Blue dashed line: Simulated amplitude of the induced spectral hole in the 980~nm profile by addressing ions within a range of 80~MHz in the 1.5~$\mathrm{\mu}$m profile. }
\end{figure}

We sweep the telecom DFB laser frequency back and forth over a 200~MHz range around the center of the $Z_1 \leftrightarrow Y_1$ absorption line during 20~ms. This generates an almost-completely-saturated spectral hole over this frequency range. A few~$\mu$s later, the inhomogeneous line of the $Z_1 \leftrightarrow X_1$ transition is scanned in 0.5~ms as shown in Fig.~\ref{fig:fig6}(a). Fig.~\ref{fig:fig6}(c) shows the absorption profiles at 980~nm with and without the pump at 1536~nm ($\alpha L(\nu)_{on}$ and $\alpha L(\nu)_{off}$). We visualize on the difference $\Delta (\alpha L)$ the emergence of an induced spectral hole, significantly narrower than the inhomogeneous linewidth of 850~MHz, and not a global depletion of the absorption profile. This suggests a partial correlation between the inhomogeneous distributions of the two transitions involved. The induced spectral hole can be adjusted by a Lorentzian function $A_c\mathcal{L}(\nu-\delta_c,\Gamma_c)$ with $A_c$ the amplitude, $\delta_c$ the central frequency and $\Gamma_c$ the FWHM, as illustrated by the green curve in Fig.~\ref{fig:fig6}(b) and as adjusted in Fig.~\ref{fig:fig6}(c).
We now examine how the $\Delta (\alpha L)$ profile evolves when the frequency of the pump excitation is slowly varied through the inhomogeneous absorption profile at 1536~nm. For this, we resort to a quasi-continuous sequence (see Fig.~\ref{fig:fig7}(a)) where the pump and readout are performed simultaneously, with the pump slowly scanned over 7~GHz in 5~s. Given the population lifetime of the ions in $Y_1$ (around 10~ms), this is equivalent to sequentially exciting ions over frequency intervals on the order of a few tens of MHz. Simultaneously, the entire inhomogeneous absorption profile at 980~nm is readout every 5~ms. This method enables us to link each 980~nm absorption profile with the hole position at 1536~nm, and obtain a map of $\Delta (\alpha L)$ in Fig.~\ref{fig:fig7}(b). 
For each value of the initial hole position, we fit the induced spectral hole $\Delta(\alpha L)$ with the Lorentzian function $A_c\mathcal{L}(\nu-\delta_c,\Gamma_c)$, and display the resulting fit parameters ($\Gamma_c$, $\delta_c$, $A_c$) in Fig.~\ref{fig:fig7}(c). We observe that the central frequency of the induced spectral hole $\delta_c$ shifts monotonously with the initial hole frequency. This results confirms the presence of partial correlation between the inhomogeneous distributions of the two optical transitions, since the induced spectral hole is always broader than the initial spectral hole, which should be a few tens of MHz wide. The minimal spectral hole width (537~MHz) is obtained when the burning frequency is tuned to the center of the inhomogeneous absorption line and becomes about 3~times broader on the edges. Therefore the correlations are not uniform across the entire inhomogeneous profile: ions located at the edges of the profile -- which likely experience stronger crystal field perturbations -- exhibit weaker correlations. We conclude that the correlations between transitions are influenced by the local crystal field environment.

\subsection{\label{sec:simplemodel}Simple model for correlation measurement}
We design a statistical approach to model the correlated absorption profiles of the 1.5~$\mu$m and 980~nm transitions. We build the 1.5~$\mu$m inhomogeneous absorption profile by summing $2\times10^5$ individual spectral lines. Each line corresponds to a symbolic ion and is assumed to have zero homogeneous width, which is justified since the homogeneous linewidth of erbium ions ($\sim$~MHz \cite{bottger_optical_2006}) is negligible compared to the inhomogeneous broadening ($\Gamma_{\mathrm{inh}}\sim1$~GHz). Within this framework, each ion at 1.5~$\mu$m, characterized by a position $\nu$, is assigned a position $\nu'$ in the 980~nm profile. This position $\nu'$ is drawn from a distribution defined by the experimentally determined parameters $\Gamma_c(\nu)$ and $\delta_c(\nu)$ (red dashed lines of Fig.~\ref{fig:fig7}(c)). The resulting inhomogeneous profile at 980~nm is well described by a Lorentzian profile with a linewidth of $0.860(0.002)$~GHz, in good agreement with the experimental value of 0.85~GHz (see Appendix~D for details). Furthermore, the simulation enables the prediction of the induced hole amplitude $A_c$ (blue dashed line in Fig.~\ref{fig:fig7}(c)) assuming a 80~MHz-wide hole burnt at 1536~nm. Overall, the statistical model appears to capture the main aspects of the observed behavior, and will be used to show consistency of our results.

\subsection{\label{sec:pumping980}Pumping at 980~nm, readout at 1536~nm}
\begin{figure}[t!]
\includegraphics[width=0.8\linewidth]{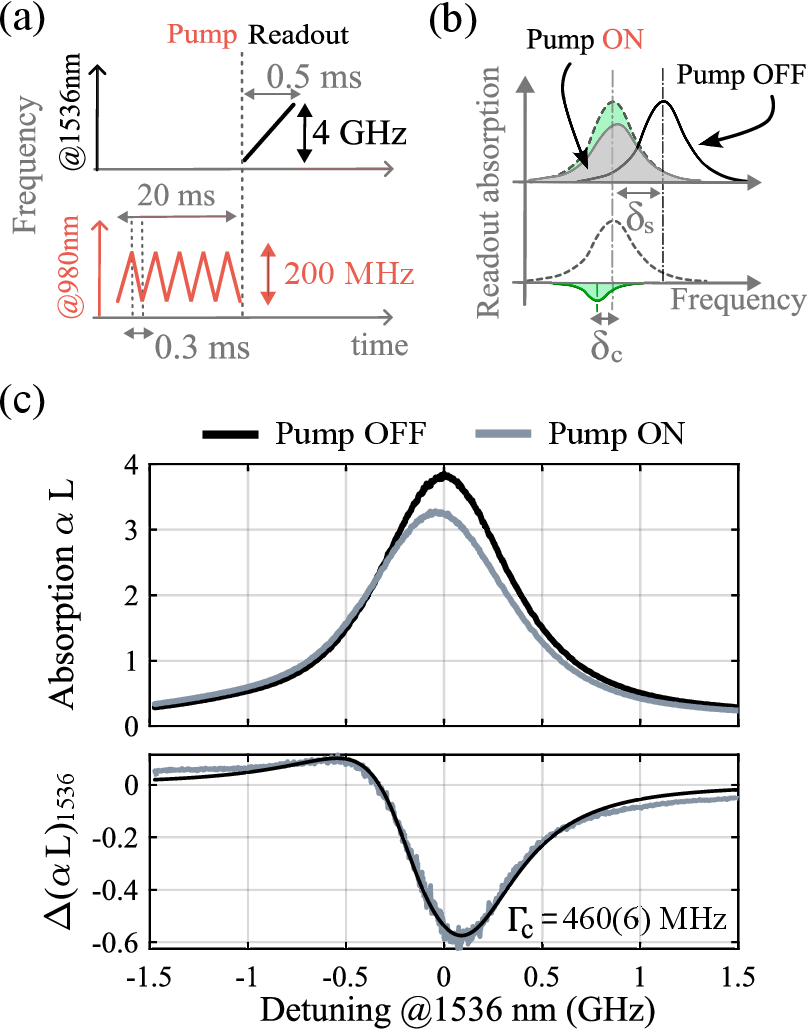}
\caption{\label{fig:fig8} (a) Chronogram. (b) Schematic absorption spectra of the probed inhomogeneous profile with and without pump detuned by $\delta_s$ (top), highlighting the induced spectral hole detuned by $\delta_c$ (bottom, green). (c) Absorption spectra of the 1536~nm transition (top) with the resulting induced spectral hole $\Delta(\alpha L)$ (bottom), for 75~ppm-Er$^{3+}$:YSO. The extracted parameters using equation 2 are: $\delta_s =60.2(0.1)$~MHz and $\delta_c = 169(2)$~MHz.
 }
\end{figure}

In this section, as a complementary approach to the previous study, we investigate the possible correlation between the two inhomogeneous distributions by interchanging the pumping and readout wavelengths (980~nm and 1536~nm, respectively). An initial spectral hole is prepared in the $Z_1 \leftrightarrow X_1$ line, followed by a readout of the whole $Z_1 \leftrightarrow Y_1$ transition (see Fig.~\ref{fig:fig8}(a)). As in the previous configuration (Section~\ref{sec:pumping1536}), we observe an induced spectral hole in the 1.5~$\mu$m line (see Fig.~\ref{fig:fig8}(b)), indicating correlated inhomogeneous distributions between the two absorption profiles. Again, the induced spectral hole is significantly broader (460~MHz) than the initial spectral hole (200~MHz), suggesting only partial correlation. However, this time $\Delta (\alpha L)$ is no longer Lorentzian and exhibits both positive and negative regions. This can be explained by the emergence of an overall shift of the inhomogeneous profile, as shown in Fig.~\ref{fig:fig8}(c). \\

\begin{figure}[b!]
\includegraphics[width=0.8\linewidth]{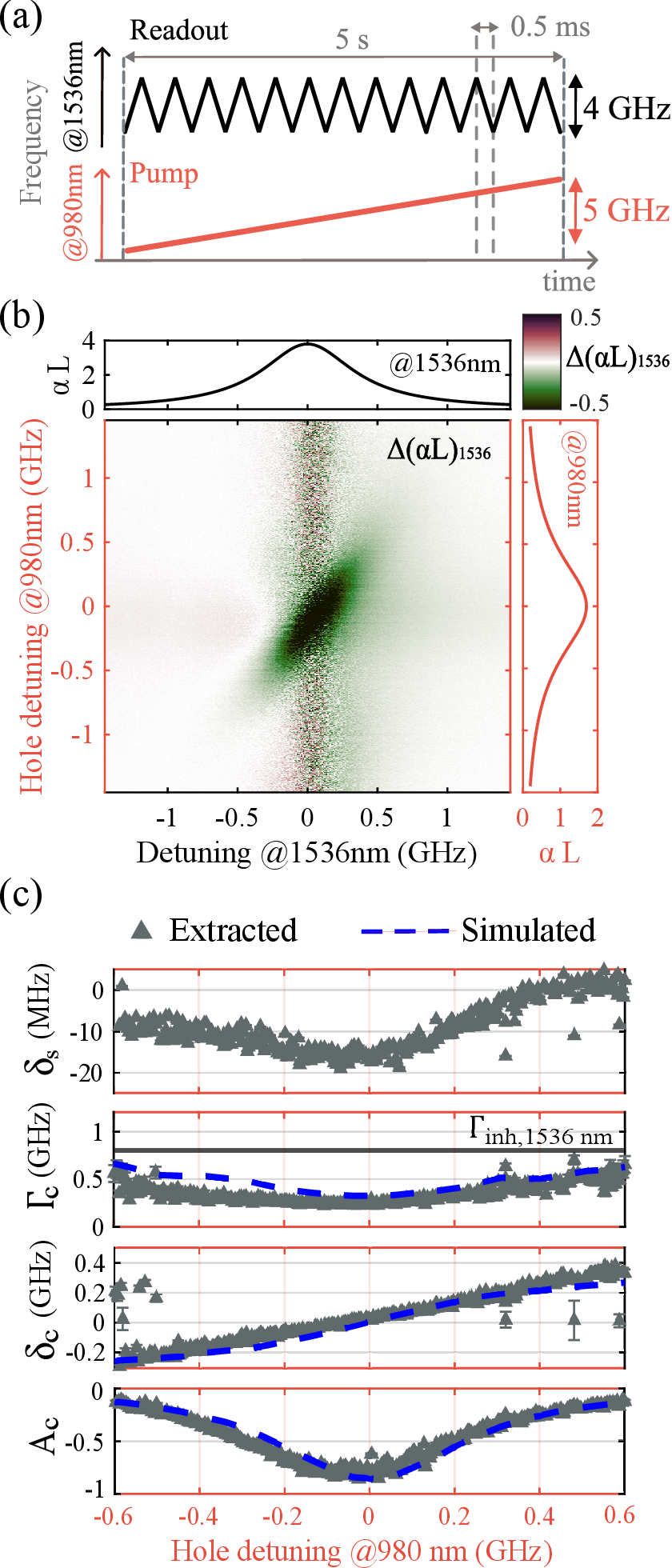}
\caption{\label{fig:fig9} (a) Experiment chronogram. (b) Colorplot of $\Delta(\alpha L)$, with the initial absorption profiles shown in side panels, for 75~ppm-Er$^{3+}$:YSO. (c) Grey triangles: extracted parameters of $\Delta(\alpha L)$ profiles using Equation (2). Blue lines: simulated parameters obtained from the modelling of the inhomogeneous profile at 980~nm, based on correlation parameters presented in Section~\ref{sec:simplemodel} and for a 120~MHz-wide initial spectral hole.
 }
\end{figure}

As in Section~\ref{sec:pumping1536}, we examine how the $\Delta (\alpha L)$ profile at 1536~nm changes with the position of the spectral hole at 980~nm, by simultaneous pumping and readout (see Fig.~\ref{fig:fig9}(a)). We represent $\Delta (\alpha L)$ as a colorplot in Fig.~\ref{fig:fig9}(b). The vertical feature at zero frequency detuning is attributed to detection noise due to the high absorption at the center of the 1536~nm line (see Fig. 12 in Appendix D). As in Section~\ref{sec:pumping1536}, the $\Delta (\alpha L)$ profiles exhibit a diagonal trend, but this time with a positive part on the left, corresponding to an excess of absorption, compatible with an overall shift of the inhomogeneous profile triggered by the pumping beam. A physical explanation for this shift is proposed in Section~\ref{sec:discussion}. We adjust $\Delta (\alpha L)$ with a sum of three Lorentzian $\mathcal{L}$ terms: 
\begin{equation}
\Delta (\alpha L) = A \mathcal{L} (\nu - \delta_s, \Gamma_{inh}) + A_c \mathcal{L}(\nu - \delta'_c, \Gamma_c) - A \mathcal{L} (\nu, \Gamma_{inh})
\end{equation}

where the first term corresponds to the absorption profile at 1536~nm with the pump ON, shifted by $\delta_s$ (green profile on Fig.~\ref{fig:fig8}(b)). The second term describes the induced spectral hole, shifted by $\delta'_c=\delta_s+\delta_c$ and characterized by a linewidth $\Gamma_c$. The third term represents the original inhomogeneous absorption profile, characterized by a linewidth $\Gamma_{inh}$. The experimental profiles $\Delta (\alpha L)$ are fitted using Equation (2), with the resulting fit parameters ($\delta_s$, $\Gamma_c$, $\delta_c$, and $A_c$) shown in Fig.~\ref{fig:fig9}(c). We observe that the induced spectral hole position shifts monotonously with the initial hole position in the 980~nm line, as in the first configuration, but this time with a 0.55 proportionality coefficient. The induced spectral hole is again narrowest (250~MHz) when the initial hole is burnt at the center of the 1536~nm inhomogeneous profile, which is more than twice as narrow as the equivalent in the other configuration. We also see that the overall lineshift $\delta_s$ reaches its absolute maximum value ($-15$~MHz) at this point.

In Section~\ref{sec:simplemodel}, using the correlation parameters derived from pumping at 1536~nm and readout at 980~nm, we were able to reconstruct the complete absorption profile at 980~nm. Based on the resulting distributions at 1536~nm and 980~nm, we can now analyze the scenario of pumping at 980~nm and probing at 1.5~$\mu$m to extract the correlation parameters $\delta_c$, $\Gamma_c$ and $A_c$. Considering a 120~MHz-wide spectral hole burned at 980 nm, the simulated values of these parameters are shown in Fig.~\ref{fig:fig9}(c). We find a strong consistency between the experimentally extracted parameters and the simulated position $\delta_c$, linewidth $\Gamma_c$ and amplitude $A_c$ of the induced hole. These results demonstrate the efficiency of our simple model, which allows for the retrieval of correlation parameter variations within the inhomogeneous profiles of both transitions, based only on the measurement of correlation in a single pumping scheme.

\section{\label{sec:discussion}Discussion}
With our non-standard spectral holeburning experiments, where a spectral hole is prepared on one transition and the readout is achieved on another transition sharing a ground state,  we were able to demonstrate the emergence of an induced spectral hole, significantly broader than the initial hole. We observed that the distribution of excited ions is broader in the edge of the readout absorption profile, as evidenced by an increased hole width $\Gamma_c$.

In the hypothetical case of total correlation, one would expect the linewidth $\Gamma_c$ to be small (of the order of the initial spectral hole) and constant with respect to spectral hole detuning. The central position $\delta_c$ should then shift with a slope equal to the ratio of the inhomogeneous linewidths (1.07 for pumping at 1536~nm and 0.94 at 980~nm). However, our measurements reveal lower slope coefficients, 0.8 and 0.55, respectively. This deviation can be explained by the fact that $\Gamma_c$ is not constant but instead increases significantly with the initial hole detuning. In addition, $\Gamma_c$ differs in magnitude for the two distributions, indicating that the ions are subject to different broadening mechanisms on each transition. These observations provide new insights into the underlying inhomogeneous broadening processes. Combining these experimental findings with theoretical considerations of crystal field effects would be valuable for a deeper understanding of the mechanisms at play.

Furthermore, our simulations confirm that the correlation parameters extracted from one distribution can be used to retrieve those of the other, since both distributions originate from the same ion ensemble. They also confirm the intuitive argument that total correlation between transitions is only possible when both transitions experience an identical and unperturbed environment. The investigation of such correlations between inhomogeneous absorption profiles could be further extended to matrices with higher symmetry, such as YVO$_4$ or CaWO$_4$, in which the correlations could be stronger. It would also be useful to explore these phenomena in other rare-earth ions to fully understand the processes at play, particularly the role of the hybrid character of the transitions and the shared spin and orbital angular momentum. 



Interestingly, excitation at 980~nm into the $X_1$ level induces a global shift ($\delta_s$) in the $Z_1 \leftrightarrow Y_1$ transition. The magnitude of this shift increases with the population of pumped ions, indicating that it is related to the excited state population. This might suggest the presence of instantaneous spectral diffusion (ISD). However, we rule out this explanation by arguing that in both cases the ions are pumped in $Y_1$: the ISD contribution should then be comparable in both configurations. An alternative explanation for the observed shift is a local heating of the crystal induced by non-radiative relaxation processes, which occur more massively in the configuration where the ions transit through the $X_1$ level. Indeed, while the $Y_1\rightarrow Z_1$ decay is fully radiative \cite{bottger_optical_2006}, the $X_1\rightarrow Y_N$ decay is almost exclusively non-radiative and therefore a significant phonon source. This local heating could induce a thermal gradient, leading to local stress, in turn leading to a global lineshift due to the piezospectroscopic effect \cite{louchet-chauvet_piezospectroscopic_2019}.

We believe the existence of strong correlations between optical transitions could be extremely interesting for many classical or quantum information processing applications using RE doped crystals. For example, in wideband radiofrequency spectral analysis using the photographic architecture \cite{louchet-chauvet_telecom_2020}, the writing and reading steps could be performed on two different transitions and simultaneously, thus enabling a 100\% probability of intercept with fully overlapping beams while keeping them easily separable with a dichroic miror, optimizing the dynamic range of signal detection. In the frame of long distance quantum communications, one can also consider implementing quantum storage (i.e using AFC spin wave protocol \cite{jobez_cavity-enhanced_2014}) using two-color engineering, where the preparation of the memory pulses would be at a different wavelength from the incoming photons to be stored. This could prevent fluorescence noise generated by the preparation step or overheating of the system when optical pumping is performed over a wide frequency range (i.e for spectral multiplexing). One could also imagine storing the photon on one optical transition and transferring the coherence back on a second transition after the spin wave storage step, thereby combining quantum storage with frequency conversion.

\section{\label{sec:level1}conclusion }
In this paper, we studied the spectroscopic properties of the little-known $(\mathrm{^4I_{15/2}}:Z_1 \leftrightarrow \mathrm{^4I_{11/2}}:X_1)$ transition in Er$^{3+}$:YSO at 3.5 K. We identified two separate lines at 980.68~nm and 982.00~nm that we respectively ascribed to site 1 and site 2. Both lines were characterized in terms of inhomogeneous linewidth, polarization dependence, spectral hole lifetime, magnetic response, and optical coherence lifetime, revealing strong similarities with the well-known $(\mathrm{^4I_{15/2}}:Z_1 \leftrightarrow \mathrm{^4I_{13/2}}:Y_1)$ transition in the telecom domain. 
Focusing on site 1, we revealed partial correlations between the inhomogeneous distributions of the two transitions. We also observed that the induced spectral hole became broader at the edges of the lines, likely due to increased perturbation from the crystal field. These findings enabled us to model the distribution of each spectral line by characterizing the induced spectral holes, which showed good agreement with our experimental observations. When pumping at 980~nm we observe an overall shift of the line at 1.5~$\mu$m that we attribute to local thermal stress. 
Overall, our results highlight complex interactions between the local environment and these transitions in Er$^{3+}$:YSO, governed by the host lattice and the dopant concentration. These insights into spectral broadening and cross-transition correlation are essential for advancing the deep understanding of rare-earth-doped materials and their applications in quantum information processing and photonic technologies.

\begin{acknowledgments}
The authors thank Philippe Goldner and Thierry Chanelière for fruitful discussions. The authors acknowledge the support and funding provided by the French Direction générale de l'armement (DGA) and of the French Agence Nationale de la Recherche (ANR), under grant ANR-24-CE47-1190 (project CHORIZO).
\end{acknowledgments}

\appendix

\section{}
\begin{figure}[h]
\includegraphics[width=0.8\linewidth]{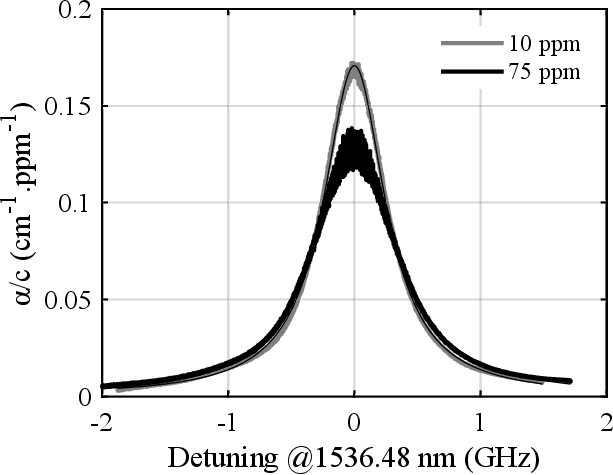}
\caption{\label{fig:fig10} Absorption profiles of site 1 of Er$^{3+}$:YSO for the $Z_1 \leftrightarrow Y_1$ transition, measured at doping concentration of 10~ppm and 75~ppm. The polarisation is along the $D_2$ axis. }
\end{figure}

Figure~\ref{fig:fig10} shows the two absorption lines at 1536.48~nm measured at the two doping concentrations, 10~ppm and 75~ppm. Both lines exhibit a Lorentzian lineshape, with the corresponding FWHM values summarized in Table~\ref{tab:table1}.

\section{}
Table~\ref{tab:table3} shows the fitted parameters for the absorption line area as a function of the polarization angle $\theta$ in the $D_1,D_2$ plane, based on the results in Fig.~\ref{fig:fig2}(b).

\begin{table}[h]
\begin{tabular}{cccc}
\hline
Experimental set & $d_1$&$d_2$&$\theta_0(\degree)$ \\
\colrule
Site 1  &1.130(0.006)&	1.410(0.006)&	-4(1) \\
 Site 2  &0.150(0.002)&	0.300(0.002)&	10.0(0.7)  \\
 \hline
\end{tabular}
\caption{\label{tab:table3} Fitting parameters of the absorption cross-section $d(\theta)$. }
\end{table}

\section{}
At low magnetic field ($|\mathbf{B}| < 20~$mT), we measure $\Delta_{Z_1}-\Delta_{Y_1}$ as the separation between the main and strongest side holes in SHB spectra for the $Z_1 \leftrightarrow Y_1$ transition, as shown in Fig.~\ref{fig:fig3}. Since the effective $g$-factors $g_{Z_1, \mathrm{eff}}$ and $g_{Y_1, \mathrm{eff}}$ are known \cite{sun_magnetic_2008} (see Table~\ref{tab:table2}), this splitting allows us to accurately determine the applied magnetic field strength.
\begin{equation}
|\mathbf{B}| = \frac{h}{\mu_B} \frac{\Delta_{Z_1} - \Delta_{Y_1}}{g_{Z_1, \text{eff}} - g_{Y_1, \text{eff}}}
\end{equation}

The next step consists in deriving the effective $g$-factor for the $X_1$ state $g_{X_1, \mathrm{eff}}$ from the splitting $\Delta_{Z_1}-\Delta_{X_1}$ observed on the $Z_1 \leftrightarrow X_1$ transition. 

\begin{figure}[t!]
\includegraphics[width=0.8\linewidth]{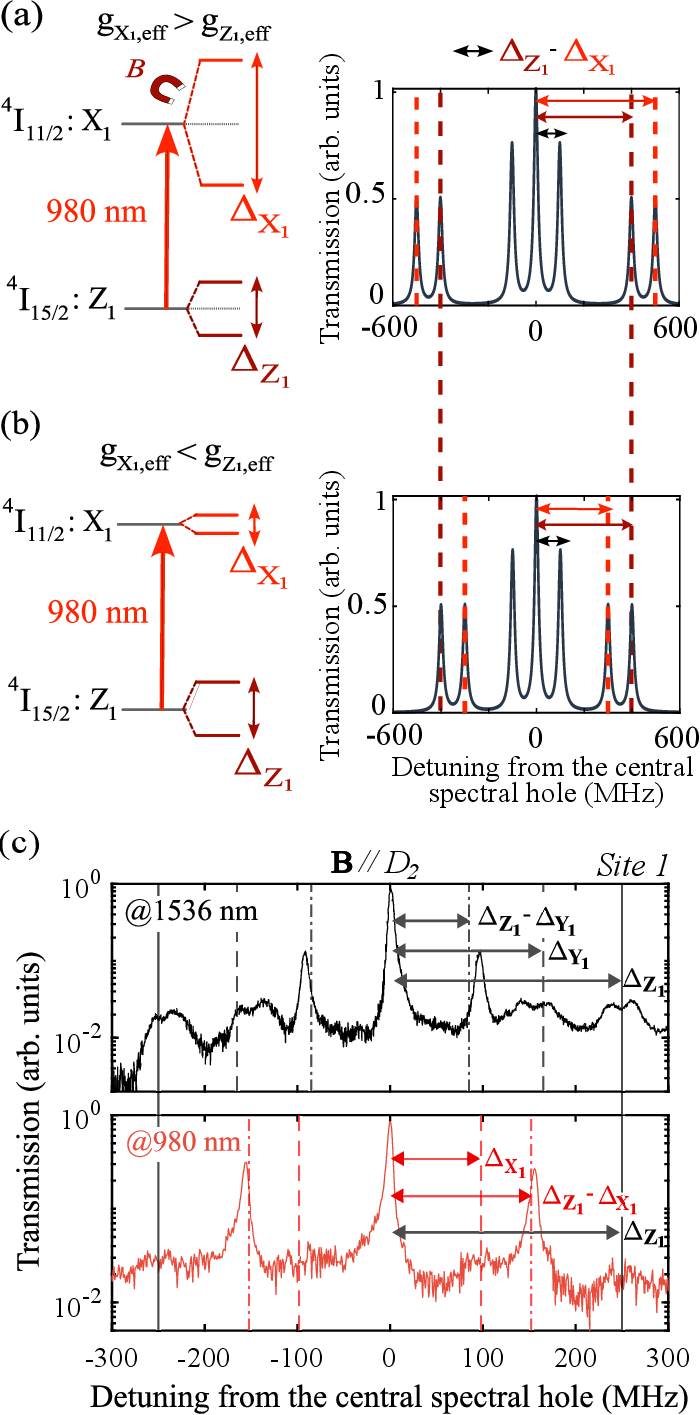}
\caption{\label{fig:fig11} (a) Energy level diagram illustrating the two configurations, depending on whether $g_{X_1, \mathrm{eff}}$ is greater or less than $g_{Z_1, \mathrm{eff}}$. (b) Resulting transmission spectra for hole burning, considering only positive detuning from the central spectral hole for simplicity. (c) Spectral hole burning transmission spectra for site 1 of 75~ppm-Er$^{3+}$:YSO under a $\sim 10$ mT magnetic field along $D_1$. The upper panel shows the $Z_1 \leftrightarrow Y_1$ transition, and the lower panel shows the $Z_1 \leftrightarrow X_1$ transition. }
\end{figure}

However, since it is not known a priori which one of $g_{Z_1, \mathrm{eff}}$ or $g_{X_1, \mathrm{eff}}$ is larger, the splitting between the central peak and the nearest side peak can be assigned to either $\Delta_{Z_1}-\Delta_{X_1}$ or $\Delta_{X_1}-\Delta_{Z_1}$. Fig.~\ref{fig:fig11}(a) and (b) show the two possible level configurations and their corresponding transmission spectra. This results in two possible solutions for $g_{X_1, \mathrm{eff}}$ :
\begin{equation}
  g_{X_1, \mathrm{eff}}= g_{Z_1, \mathrm{eff}}\pm h \frac{|\Delta_{Z_1}-\Delta_{X_1} |}{\mu_B|\mathbf{B}|}
\end{equation}

The experimental data shown in Fig.~\ref{fig:fig3} allows the unambiguous identification of the side hole at $\Delta_{Z_1}$ when $|\mathbf{B}| // D_1$. We conclude that $g_{Z_1, \mathrm{eff}}>g_{X_1, \mathrm{eff}}$. We observe the same result in Fig.~\ref{fig:fig11}(c) when $|\mathbf{B}| // D_2$ leading to :
\begin{equation}
  g_{X_1, \mathrm{eff}}= g_{Z_1, \mathrm{eff}}+ h \frac{|\Delta_{Z_1}-\Delta_{X_1} |}{\mu_B|\mathbf{B}|}
\end{equation}

\section{}
\begin{figure}[t]
\includegraphics[width=0.8\linewidth]{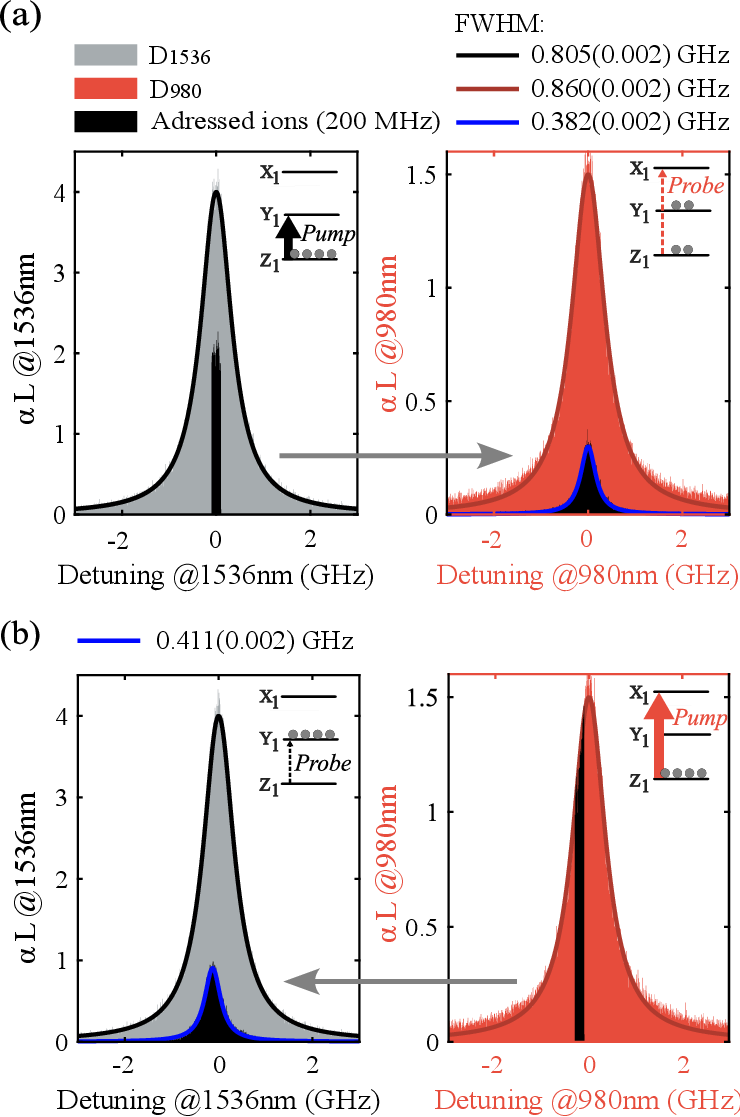}
\caption{\label{fig:fig12} (a) Simulated ion distributions at 1.5$\mu$m (left) and resulting distributions at 980~nm (right) using the correlation parameters obtained from Fig.~\ref{fig:fig7}(c). Addressed ions correspond to those within the inhomogeneous profile at 1.5$\mu$m between $-0.1$ and 0.1~GHz (black histogram). The amplitude $A_c$ is 0.3. (b) Same distributions but here the addressed ions correspond to those within the 980~nm inhomogeneous profile with frequency range between $-0.3$ and $- 0.1$~GHz. }
\end{figure}

The statistical approach used to evaluate the correlation begins by simulating the ion distribution at 1.5~$\mu$m. We consider a Lorentzian distribution of ions, denoted $D_{1536}$, composed of $N = 2 \times 10^{5}$ ions. This number of ions allows a detuning resolution of 0.1~MHz over the range from -2.5~GHz to +2.5~GHz. Each ion is assigned a detuning value $\nu_i$ (in~GHz), as illustrated in the left panel of Fig.~\ref{fig:fig12}(a).\\ For each ion $i$, we assign a value $B(\nu_i)$, which corresponds to the slope of the piecewise linear fit to the $\delta_c(\nu)$ data from Fig.~\ref{fig:fig7}(c). Additionally, a random value $C(\nu_i)$ is sampled from a Lorentzian distribution with a linewidth $\Gamma_c(\nu_i)$, where $\Gamma_c(\nu_i)$ is determined by a polynomial fit to the data presented in Fig.~\ref{fig:fig7}(c).

Subsequently, we construct a second distribution, $D_{980}$ derived from $D_{1536}$. For this, we generate, for each ion $i$, a new variable $\nu'_i$, which depends on its initial detuning $\nu_i$:
$\nu'_i=\nu_i \times B(\nu_i )+C(\nu_i)$.
The distribution $D_{980}$ corresponds to the detuning distribution in the 980~nm profile and is shown in the right panel of Fig.~\ref{fig:fig12}(a). 
We then select specific ions in the 1.5~$\mu$m profile, within a frequency range between -0.1 and 0.1~GHz, to see where these ions are located in the 980~nm profile and to directly observe the induced spectral hole (see black region in Fig.~\ref{fig:fig12}(a)). In the configuration where the pumping is achieved at 1.5~$\mu$m, considering that pumping leads to saturation of the optical transition (i.e., to equal population in the $Z_1$ and $Y_1$ levels), the number of ions probed at 980~nm is reduced by half by the pumping process. This is why we consider the addressed ions to be half of the ions that we selected within a frequency range between -0.1 and 0.1~GHz. 
By changing the frequency of the probing ions in the 1.5~$\mu$m profile, we can study how the induced spectral hole evolves as a function of the probing frequency. We fit the induced spectral hole with a Lorentzian function to extract its linewidth $\Gamma_c(\nu)$ position $\delta_c(\nu)$, and amplitude $A_c(\nu)$. 
By selecting ions in the 980~nm profile within the frequency range between -0.3 and -0.1~GHz (scenario of Fig.~\ref{fig:fig8}(b)), we can then characterize the induced spectral hole in the profile at 1.5~$\mu$m and measure the same parameters, as shown in Fig.~\ref{fig:fig12}(b). By pumping at 980nm, we can assume that we obtain population inversion in the $Y_1$ level, so the addressed ions all the ions in the pumping frequency range.
The obtained parameters are represented with blue dashed lines in Fig.~\ref{fig:fig9}(c).

\bibliography{Mabibliothequebis.bib}

\end{document}